\documentclass[twocolumn,showpacs,prb,aps]{revtex4-1}

\usepackage{graphicx}
\usepackage{amssymb}

\begin{document}

\title{VUV Brillouin scattering from superpolished vitreous silica}

\author{B. Ruffl\'e$^{1,2}$}
\author{E. Courtens$^{1,2}$}
\author{M. Foret$^{1,2}$}

\affiliation{$^1$Laboratoire Charles Coulomb, UMR 5221, Universit\'e Montpellier 2, F-34095 Montpellier, France\\
$^2$Laboratoire Charles Coulomb, UMR 5221, CNRS, F-34095 Montpellier, France}

\date{29 July 2011}

\begin{abstract}
A previous inelastic UV scattering experiment on silica glass is reproduced using a high grade superpolished sample.
In the pristine sample condition, surface scattering is not observable compared to Rayleigh scattering from the bulk.
However, exposure to a fluence of the order of 100 J/cm$^2$ at photon energies slightly below the electronic gap generates observable surface damage.
This occurs after a few hours illumination with the monochromatic spectrometer beam.
No anomaly in the Brillouin linewidth was found up to an excitation energy of 7.8 eV.

\end{abstract}

\pacs{63.50.Lm, 78.35.+c, 78.68.+m, 42.70.Ce}

\maketitle

The propagation and attenuation of hypersonic waves in glasses remains a subject of considerable interest.
It ties, among others, to the thermal conductivity anomaly, \cite{Zel71} to quasi-local vibrations and the boson peak, \cite{Par07} and to the existence of a noisy non-affine displacement field at sufficiently short scales. \cite{Leo06}
It is also a subject of active debate as it is experimentally difficult to access the crucial range of sound frequencies $\Omega$, which in vitreous silica roughly lies  at $\Omega /2\pi $ between 0.1 and 1 THz.
A choice technique is Brillouin scattering (BS) of light in which $\Omega $ increases almost linearly with the incident photon energy $E_{\rm i} = \hbar \omega _{\rm i}$.
However, the experiment becomes impossible near the electronic absorption edge of the sample, either owing to uncertainty broadening, \cite{Vac06} or more trivially due to the loss of signal. \cite{Mas06}
In vitreous silica, the edge at $E_{\rm i} \approx 8$ eV, sets an upper limit to $\Omega /2 \pi \approx 130 $ GHz in backscattering. \cite{Mas06}
Beyond the edge, the samples remain opaque to electromagnetic radiation up to soft x-rays.
BS using a near forward scattering geometry and $E_{\rm i} \approx 20$ keV gives then access to acoustic frequencies that are not much below $\approx 1$ THz. \cite{Set98,Bal10}
There is thus a gap of nearly an order of magnitude in $\Omega $ remaining inaccessible to BS.
To alleviate this problem there has been much recent work on picosecond optical techniques (POT), in particular applied to silica. \cite{Ayr11, Pon11, Kli11}.
However, POT requires the use of thin films, and although experimental scattering widths mostly follow expectations, \cite{Ayr11,Kli11} sometimes they do not.
In particular, a crossover at $\Omega /2\pi  \sim  170$ GHz was claimed \cite{Pon11} and it was stated that it might depend on film preparation. \cite{Pon11b} 
A crossover, with onset around 113 GHz, was also claimed in the inelastic UV scattering experiment on bulk silica performed with the IUVS spectrometer at Elettra in Trieste, Italy.\cite{Mas06}
That frequency corresponds to an excitation energy $E_{\rm i} \simeq 7.4$ eV, close to the electronic gap.
Structural crossovers at relatively low acoustic frequencies are somewhat surprising, as they do imply correlation lengths of the order of $\Lambda /2$, where $\Lambda $ is the acoustic wavelength.
The lengths corresponding to the above frequencies seem rather large, 50 to 70 when expressed in terms of the mean linear SiO$_2$ size.

Therefore we decided to revisit the IUVS experiment on silica at Elettra.\cite{Mas06}
Until now, there seems to have been no attempt at measuring a quasi-elastic Rayleigh peak with that spectrometer.
Rather, the strong spurious scattering originating from the sample faces is often used as convenient source of light for line up.
As a result, when the sample starts absorbing significantly the exciting radiation, the observed Brillouin peaks ride on the wings of an intense elastic signal.\cite{Mas06}
To suppress the spurious elastic signal originating from the surface, we use here a very perfect and ultra-clean superpolished sample of high grade silica.
As described below, this approach is remarkably successful.
We observe a Rayleigh signal from the bulk and the Brillouin peaks do not ride on its wings.
Further, we discover that at $E_{\rm i} \geq  7.4$ eV the radiation from the instrument damages the sample surface within a few hours.
We obtain no evidence for the crossover found in [\onlinecite{Mas06}].
However, owing to the use of a different quality of silica, and given the limited access time, linewidth results are not obtained here beyond $E_{\rm i} = 7.8$ eV, corresponding to the acoustic frequency $\Omega /2\pi  = 121$ GHz.

The principle of the IUVS spectrometer is described in [\onlinecite{Mas06}] and references therein.
An important recent improvement is the top-up operation of the Elettra synchrotron, with injection every 6 minutes.
Spectra taken at different times are thus easily compared.
Our sample is a 5 cm diameter, 5 mm thick, Suprasil 312 flat from Heraeus Quarzglas.
This variety of synthetic silica is among preferred ones for several demanding applications such as high quality-factor oscillators \cite{Age04} or optics in laser-ignition facilities requiring highly perfect finishes.\cite{Sur11}
Our sample contains about 220 ppm OH as checked by infrared transmission.
It received on both sides an {\em enhanced} superpolish consisting in alternate etching with HF and superpolishing steps.  
This removes deep surface damage, as described in [\onlinecite{Sur11}].
As final preparation step, the sample was carefully washed, rinsed, and dried in a clean room, and then sealed in a hermetic sample holder for transportation.
At Elettra, the holder was mounted on the {\em xyz}-manipulator, introduced in the sample chamber that was then flushed with dust-free air before evacuation, at which point the cover of the holder popped up and was removed exposing the surface.

The near backscattering geometry is shown in the inset of Fig. 1.
The sample normal is inclined by about 35$^\circ$ from the horizontal.
The scattering angle is $2\theta = 172^\circ$.
\begin{figure}
\includegraphics[width=8.5cm]{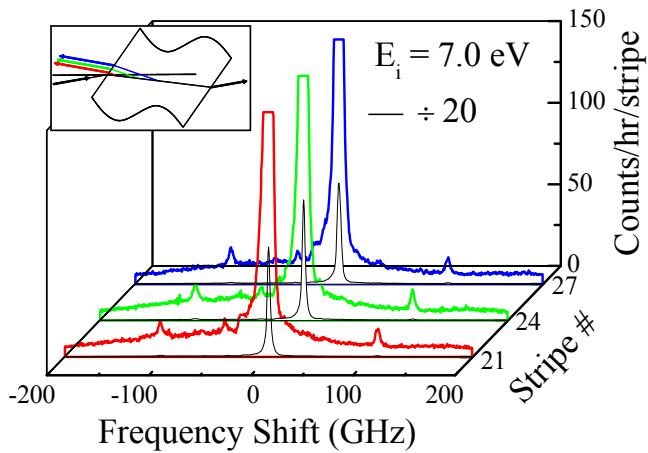}
\caption{(Color online) Rayleigh-Brillouin spectra collected at three different depths at $E_{\rm i} =  7.0$ eV. The signals result from a 4-hour average. The Brillouin peaks at $\pm 107$ GHz ride on a  slightly sloping background. The Rayleigh peaks divided by 20 are also shown. The inset illustrates the scattering geometry and the collection paths for the three spectra.} 
\end{figure}
The incident beam impinges on the sample from the bottom left.
It is inclined by an angle $\delta  = 4^\circ$ from the horizontal. 
For clarity that angle, and the symmetrically located scattering direction, are increased to 10$^\circ$ in the drawing.
The beam is polarized horizontally, perpendicular to the plane of the drawing.
Although this maximizes the Fresnel reflection, it also minimizes the diffuse surface scattering. \cite{Els75}
A magnified and dispersed image of the scattered light is collected on a CCD.
On this image, different vertical positions correspond to different depths, as illustrated.
Successive rows on the CCD are here binned by 4 to decrease the reading noise.
These binned rows are called {\em stripes} in the following, the sample depth corresponding to 16 stripes.
The spectrometer disperses the light horizontally.
Brillouin peaks are seen on successive stripes, as illustrated in Fig. 1 for $E_{\rm i} =  7.0$ eV.
The focusing on the sample consists in obtaining a sharp image of the desired scattering volume on the CCD.
To this effect a narrow strip of Al was deposited through the middle of the front face before the final cleaning.
This strip was moved in the beam with the manipulator to provide a signal for sharp focusing.
This is important as the depth of field is quite restricted.
It can be seen on Fig. 1 that the sizes of the Brillouin and Rayleigh signals are nearly constant from stripe \#21, near the front face of the sample, to stripe \#27, towards its middle.
However, the width of the resolution function clearly increases from stripe \#21 to \#27.
This loss of resolution results from defocussing.
Useful spectra for strength and linewidth determination are only obtained on five successive stripes.
These are then averaged for data treatment.

Fig. 2 illustrates the integrated Rayleigh intensity in function of depth and time at a somewhat higher energy $E_{\rm i} =  7.4$ eV.
The first hour is accumulated on a fresh location of the sample face.
\begin{figure}
\includegraphics[width=8.5cm]{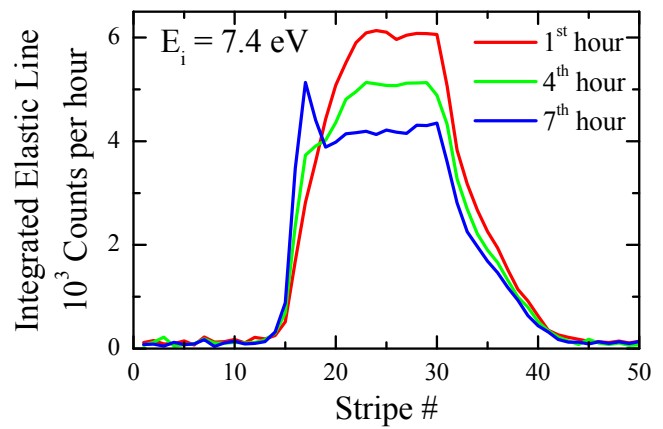}
\caption{(Color online) The integrated Rayleigh line in function of depth at $E_{\rm i} =  7.4$ eV. The first hour is collected on a pristine location of the sample surface, and the other signals illustrate the degradation of the surface in function of time.} 
\end{figure}
The front surface, at stripe \#17, generates half the peak intensity of 6 kCounts/h.
Similarly, the back face is at stripe \#33.
Moving inside to stripe \#20, the signal reaches 85\% of its peak intensity.
This slow intensity increase relates to the vertical extent of the beam, approximately equal to 40 $\mu $m at focus, coupled to the smallness of $\delta  $ which produces a long intercept between the incident and scattered directions.
Moving away from focus, the intensity variation with depth becomes even slower owing to the divergence of the incident and scattered beams.
This is observed near the exit face, with a decay extending from stripe \#30 to beyond stripe \#40.
The signal accumulated during the 4$^{\rm th}$ hour shows an excess of scattering originating from the front face, at stripe \#17, while the signal collected from the bulk decreases.
Both effects are doubled during the 7$^{\rm th}$ hour, showing that they are cumulative and linear.
Similar effects, with strength depending on $E_{\rm i}$, were observed at all energies at and above 7.4 eV.
Microscopic examination of the sample after the experiment reveals the presence of faint diffuse cloudiness on the exposed spots.
It is obvious from Fig. 2 that this damage significantly degrades the surface finish.
It scatters the incoming beam on its way in, and it scatters or defocuses the signal on its way out.
It is well known that synthetic vitreous silica exhibits various absorption bands in the region between 7 and 8 eV related {\em e.g.} to strained  Si-O-Si bonds or to hydroxyl groups $\equiv$Si-OH. \cite{Kaj07}
UV absorption in these bands can cause either compaction or expansion, as reviewed in [\onlinecite{Kue03}], and this in one-photon processes. \cite{Kaj03,Smi06}
Such damage becomes observable after an exposure of typically 100 J/cm$^2$.
The flux of the IUVS instrument is of the order of $ 3 \times 10^{12}$ photons/s, which for a waist area of  $ 2 \times 10^{-4}$ cm$^2$ and at 8 eV gives an irradiance of nearly 0.02 W/cm$^2$. 
After one hour, the beam fluence is then indeed $\sim 70$ J/cm$^2$, in agreement with our observations.
In consequence, data used for the analysis presented below were accumulated at a given spot for a maximum of 4 hours, after which the sample was systematically translated to a new pristine location.
Indeed, damage evolution in the bulk could progressively shift the Brillouin frequency and thereby jeopardize linewidth measurements.
Therefore we checked that the same results are obtained, within experimental uncertainty, whether the data are averaged for four successive hours, or for the first two hours, or for the last two hours.
Such degradation can become particularly bothersome at the highest $E_{\rm i}$ values where data accumulation takes days owing to sample absorption.

Figure 3 illustrates Brillouin peaks at three $E_{\rm i}$ values.
\begin{figure}
\includegraphics[width=8.5cm]{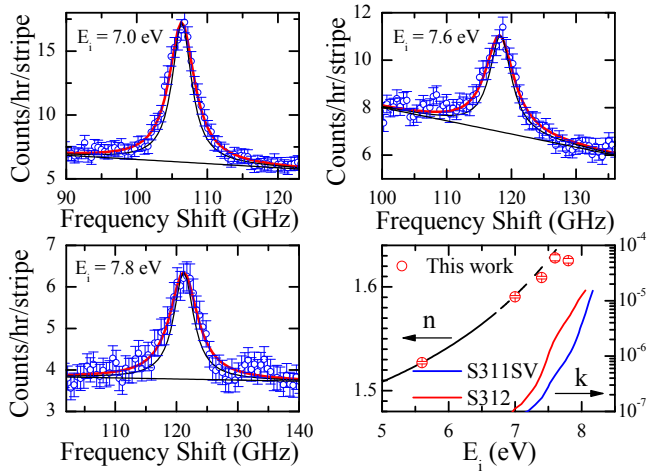}
\caption{(Color online) Brillouin peaks at three $E_{\rm i}$ values (dots). The adjustment to a DHO riding on a sloping background is shown by the thick (red) curve. The elastic line is the thinner (black) curve. The last panel shows the real ($n$) and imaginary ($k$) parts of the refractive index.} 
\end{figure}
The data are adjusted to a damped harmonic oscillator (DHO) line shape superposed to a sloping background.
The DHO of position $\Omega$ and width $\Gamma$ is convoluted with the instrumental response function derived from the simultaneously measured and averaged quasi-elastic line illustrated in the same plots.
From the values of $\Omega$ we derive the dispersion of the refractive index $n$ assuming a constant sound velocity $v = 5960$ m/s.\cite{Vac05}
To this effect we use $v = \Omega /Q$, where the scattering vector $Q$ is given by $Q = 4\pi n \sin \theta ' / \lambda $ with the vacuum wavelength $\lambda = h c / e E_{\rm i}$.
Here, $2 \theta '$ is the internal scattering angle corresponding to the external $2\theta $.
The resulting values of $n$ are shown in Fig. 3.
They agree well with the curve provided by the manufacturer,\cite{priv} except for the highest energy point where that curve was not measured but extrapolated from lower energies as shown by the dashed line. 
One should note that the point at 7.6 eV is affected by a larger error owing to the stronger slope of the background seen in Fig. 3.
This background seems to originate from a radiation halo emitted by the upstream bending magnets in the synchrotron.
Its strength sensitively depends on the exact spectrometer line-up.

The collected Brillouin intensity considerably decreases from 7.0 to 7.8 eV.
The data at 7.0 eV were obtained in only 4 hours, while these at 7.8 eV required 16 hour averaging.
Besides the reduced luminosity of the instrument as $E_{\rm i}$ increases, a strong reduction occurs at 7.8 eV owing to the sample absorption of the incident and scattered light.
The absorption coefficient is $\alpha = 4\pi k / \lambda $, where $k$ is the imaginary part of the refractive index shown in Fig. 3 where Suprasil S312 (at 250 ppm OH) is compared to a well dried variety S311SV (at 50 ppm OH).\cite{priv}
In our sample, at 7.8 eV one has $\alpha \approx 4$ cm$^{-1}$, while for the dry variety this value is only reached beyond 8 eV.
With $\alpha = 4$ cm$^{-1}$, nearly two thirds of the Brillouin signal are lost by absorption.
With $\alpha = 10.5$ cm$^{-1}$, the value for our sample at 8 eV, only about 10\% of the signal remains, which would require a ten times longer counting than at 7.8 eV to achieve a similar accuracy on $\Gamma $, making the experiment practically impossible.
It should be remarked that the absorption in S312 is well above the extrapolated Urbach tail.\cite{Sai00}
Incidentally, the uncertainty broadening \cite{Vac06} given by $\Delta \Omega = 2k \Omega /n $ remains totally negligible at 7.8 eV, with $\Delta \Omega /2 \pi $ below 1 MHz.

Fig. 4 compares the fitted values of our full widths $\Gamma /2\pi $ to the previous IUVS results\cite{Mas06} and to POT data\cite{Ayr11,Kli11}.
\begin{figure}
\includegraphics[width=8.5cm]{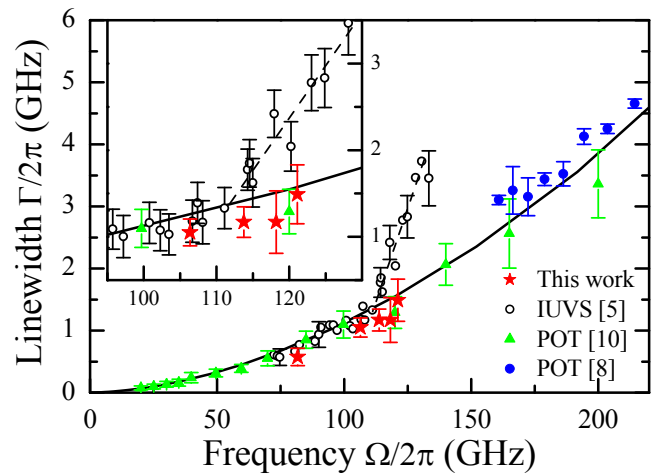}
\caption{(Color online) Brillouin full widths compared to the previous IUVS result and to recent POT data, the full line being the theoretical expectation explained in the text. The dashed line is a guide to the eye. The inset is an enlargement of the central region including all error bars.} 
\end{figure}
Also included is the prediction based on thermally activated relaxation plus anharmonicity\cite{Vac05} to which a contribution in $\Omega ^4$ owing to quasi-local vibrations is added as discussed 
in [\onlinecite{Ayr11}].
While the previous IUVS results \cite{Mas06} strongly depart from the predicted full line, as shown by the dashed line, the new results follow normal expectations over their narrow range.
They also line up with the POT results although these are obtained on various qualities of thin film silica, wet thermal oxidized at 1050 $^\circ$C in [\onlinecite{Ayr11}] while produced by plasma enhanced chemical vapor deposition, a much lower temperature process, in [\onlinecite{Kli11}].
The inset emphasizes that we do not find here a crossover at 113 GHz.
It is clear from the above that IUVS is a delicate technique.
The determination of linewidths could be affected {\em e.g.} by the wing of the Rayleigh line or by accumulating damage of the sample, among other artifacts.
However, it would be preposterous to conclude at this stage that the {\em measurement} in [\onlinecite{Mas06}] is necessarily in error.
It might also occur that the values are correct but should not be interpreted in terms of an anomalous increase in damping.
This could {\em e.g.} be the case if additional refractive index fluctuations would occur close to the absorption edge of the glass producing an enhanced {\em uncertainty broadening} in BS.

For that reason we also attempted extracting the Landau-Placzek ratio $R_{\rm LP} = I_{\rm R}/2I_{\rm B}$, where $I_{\rm R}$ is the integrated Rayleigh intensity and $2I_{\rm B}$ the integrated Brillouin doublet intensity.\cite{Lan34}
In silica at room temperature, $R_{\rm LP}$ should be of the order of 20 to 25,\cite{Buc74,Kro86} and no unusual increase in Rayleigh scattering is anticipated up to 6.3 eV.\cite{Sch06}
Instead, we found much higher values of $R_{\rm LP}$, about twice the expected value at 5.6 eV, and trice at 7.8 eV.
However, we observed that the depth profile of the Rayleigh signal at 7.8 eV is very similar to that at 7.4 eV shown in Fig. 2.
Owing to the relatively strong sample absorption at 7.8 eV we would have expected a triangular shape, almost zero near the exit face.
It suggests that a very significant part of $I_{\rm R}$ is produced by a spectral pollution of longer wavelength which contributes to $I_{\rm R}$ but not to the Brillouin signal.
In such conditions, a significant $R_{\rm LP}$ is not obtained.
It would be of considerable interest to resolve that issue.
Indeed, a measurement of $R_{\rm LP}$ in the Urbach tail might reveal the presence of clusters relevant to the electronic gap.\cite{Ina10}

We have shown in this work that it is possible in IUVS from silica to reduce surface scattering to a quantity that is negligible compared to bulk Rayleigh scattering.
This is achieved by using an extremely high quality surface finish combined with an appropriate sample handling procedure.
A main finding is that the IUVS light beam, working at energies slightly below the absorption edge of silica, is then sufficient to damage the sample surface within a few hours.
Although the origin of damage depends on sample composition, it also occurs in OH-free silica.\cite{Kaj03}
This was presumably masked in the experiment reported in [\onlinecite{Mas06}] owing to strong surface scattering.

The authors express many thanks to Drs. H. Bercegol and Ph. Bouchut from the Commissariat \`{a} l'\'{E}nergie Atomique, France, for the exceptional sample and its final cleaning, to J. Barbat from the Laboratoire Charles Coulomb, Montpellier, for the realization of the hermetic sample holder, to Dr. Bodo K\"{u}hn from Leybold Heraeus, Hanau, for data on Suprasil, and to Dr. C. Masciovecchio from Elettra, Trieste, and Dr. R. Vacher from Laboratoire Charles Coulomb, Montpellier, for valuable remarks.
Last but not least, the authors warmly thank Dr. Silvia Di Fonzo and Alessandro Gessini from Elettra who participated as local contacts in two successive measurement runs and whose names where included in the author list of an early draft.
Without their support the experiment would not have been possible.

\end{document}